\documentclass[12pt]{article}
\usepackage[english]{babel}
\usepackage{amsmath}
\usepackage{latexsym}
\usepackage{amssymb}
\usepackage[all]{xy}
\usepackage[dvips]{graphicx}
\usepackage{epsfig}
\usepackage{psfrag}

\numberwithin{equation}{section} \numberwithin{table}{section}
\numberwithin{figure}{section}

\begin{document}


\begin{titlepage}
  \begin{flushright}
  {\small CQUeST-2010-0349}
  \end{flushright}

  \begin{center}
  \vspace{20mm}

  {\LARGE \bf Holography of Charged Dilaton Black Holes in General Dimensions}

  \vspace{10mm}

  Chiang-Mei Chen$^1$ and Da-Wei Pang$^2$

  \vspace{5mm}

  {\small \sl ${}^1$ Department of Physics and Center for Mathematics and Theoretical Physics,\\
                   National Central University, Chungli 320, Taiwan}

  {\small \sl ${}^2$ Center for Quantum Spacetime, Sogang University, \\
                     Seoul 121-742, Korea}

  {\small \tt cmchen@phy.ncu.edu.tw, \quad pangdw@sogang.ac.kr}

  \vspace{10mm}

  \end{center}

\begin{abstract}\baselineskip=18pt
We study several aspects of charged dilaton black holes with planar symmetry in
$(d+2)$-dimensional spacetime, generalizing the four-dimensional results
investigated in arXiv:0911.3586 [hep-th]. We revisit the exact solutions
with both zero and finite temperature and discuss the thermodynamics
of the near-extremal black holes. We calculate the AC conductivity in the
zero-temperature background by solving the corresponding Schr\"{o}dinger
equation and find that the AC conductivity behaves like $\omega^{\delta}$,
where the exponent $\delta$ is determined by the dilaton coupling $\alpha$ and
the spacetime dimension parameter $d$.
Moreover, we also study the Gauss-Bonnet corrections to $\eta/s$ in a 
five-dimensional finite-temperature background.
\end{abstract}

\setcounter{page}{0}

\end{titlepage}

\pagestyle{plain} \baselineskip=19pt

\tableofcontents

\section{Introduction}
The AdS/CFT correspondence~\cite{Maldacena:1997re, Aharony:1999ti}
has been shown to be a useful tool for studying the dynamics
of a strongly coupled field theory, since many physical properties
of field theory can be derived by its weakly coupled and
tractable dual gravitational description.
In recent years numerous applications for the AdS/CFT correspondence, such
as in QCD etc., have been explored in detail.
In particular, the investigation of certain condensed matter systems,
namely the AdS/CMP correspondence, has accelerated quickly in the past years.
Some excellent reviews have appeared~\cite{Hartnoll:2009sz}.

In order to study the gravity dual of a condensed matter system
at finite temperature, we need to consider a suitably corresponding black hole
for the background spacetimes.
One particular class of interesting backgrounds is the charged dilaton black
holes~\cite{Gibbons:1987ps, Preskill:1991tb, Garfinkle:1990qj, Holzhey:1991bx}
of the Einstein-Maxwell-dilaton theory in which the dilaton field is
exponentially coupled with the gauge field, $\mathrm{e}^{2 \alpha \phi} F^2$.
A specific property of this type of black hole is that the Bekenstein-Hawking
entropy vanishes at the extremal limit for any value of the non-zero dilaton coupling,
$\alpha \neq 0$, therefore the higher curvature corrections are crucial.
In addition, the charged dilaton black holes with a Liouville potential were studied, e.g.
in~\cite{Cai:1996eg}, and the results suggest that their AdS generalizations may provide
interesting holographic descriptions of certain condensed matter systems.

Recently, the holography of charged dilaton black holes in AdS$_4$ with planar symmetry
was extensively investigated in~\cite{Goldstein:2009cv}.
It turns out that the near horizon geometry was Lifshitz-like with a dynamical
exponent $z$ determined by the dilaton coupling. The global solution
was constructed via numerical methods, and the attractor behavior was
also discussed. The authors also examined the thermodynamics of near
extremal black holes and computed the AC conductivity in a
zero-temperature background. For related work on charged dilaton
black holes see~\cite{Gubser:2009qt, Gauntlett:2009bh,
Cadoni:2009xm, Charmousis:2010zz}.

In this paper we generalize the work of~\cite{Goldstein:2009cv}
in four dimensional spacetime to arbitrary $(d + 2)$-dimensions.
For a practical application to a specific system in condensed matter physics,
the value of $d$ is given. For example, one should choose $d = 2$ to study the
(2 + 1)-dimensional layered systems.
However, our physical spacetime may be higher dimensional with tiny extra
dimensions, and more spacetime dimensions might be holographically generated
when extra adjoint fields are involved.
Therefore, it is of interest to consider the generalization to various dimensions
and try to find the universal behavior.

By considering a $(d + 2)$-dimensional Einstein-Maxwell-dilaton action with dilaton
coupling of the form $\mathrm{e}^{2\alpha\phi}$, in both zero and finite
temperatures, we obtain particular exact scaling solutions which are expected to be the
near horizon geometries of the considered black hole backgrounds in
AdS$_{d+2}$. The zero-temperature solution is still Lifshitz-like
and the dynamical exponent is determined by $\alpha$ and the
spacetime dimension. The thermodynamics of the near extremal black
holes is also studied. Furthermore, we compute the AC conductivity
in the zero-temperature background. We can transform the
corresponding equation of motion into a Schr\"{o}dinger form with an
effective potential of the form $V(z) = c/z^2$ and then determine the
frequency dependence of the AC conductivity as $\mathrm{Re}(\sigma) \sim
\omega^\delta$. Here both constants $c$ and $\delta$ are
determined by $\alpha$ and $d$. Moreover, we compute the
Gauss-Bonnet corrections to $\eta/s$ in a five-dimensional finite
temperature background, and the result agrees with the well-known AdS
counterpart when the dynamical exponent $z \to 1$.

The rest of the paper is organized as follows. In Section 2 we
present the exact solutions in $(d + 2)$-dimensional spacetime.
The thermodynamics of the near extremal black holes is discussed in
Section 3, and the AC conductivity is calculated in Section 4. The derivation of
the Gauss-Bonnet corrections to $\eta/s$ in a five-dimensional finite temperature background is
given in Section 5. A summary and discussion will be given in the final part.

\section{The solution}
In this section we will exhibit the exact solutions, including both zero and
finite temperature cases. Consider the following Einstein-Maxwell-dilaton action in
$(d+2)$-dimensional spacetime with a negative cosmological constant:
\begin{equation} \label{2eq1}
S = \frac1{16 \pi G_{d+2}} \int d^{d+2}x \sqrt{-g} \left( R - 2 \Lambda - 2
\partial_\mu \phi \partial^\mu \phi - \mathrm{e}^{2 \alpha \phi}
F_{\mu\nu} F^{\mu\nu} \right),
\end{equation}
the corresponding equations of motion can be summarized as follows:
\begin{eqnarray}
&& R_{\mu\nu} - \frac12 R g_{\mu\nu} + \Lambda g_{\mu\nu}
= 2 \partial_\mu \phi \partial_\nu \phi - g_{\mu\nu} (\partial \phi)^2
+ 2 \mathrm{e}^{2 \alpha \phi} F_{\mu\lambda} F_\nu{}^\lambda -
\frac12 \mathrm{e}^{2 \alpha \phi} g_{\mu\nu} F^2,
\nonumber\\
&& \partial_\mu \left( \sqrt{-g} \partial^\mu \phi \right)
= \frac{\alpha}{2} \sqrt{-g} \mathrm{e}^{2 \alpha \phi} F^2, \qquad
\partial_\mu \left( \sqrt{-g} \mathrm{e}^{2 \alpha \phi} F^{\mu\nu} \right) = 0.
\end{eqnarray}
The cosmological constant is related to the AdS radius $L$ by
\begin{equation}
\Lambda = - \frac{d (d+1)}{2 L^2},
\end{equation}
and we will set $L = 1$ for simplicity in the following.
The dependence on $L$ can be recovered simply by a dimensional analysis.

Let us take the following ansatz for the planar symmetric metric and gauge potential:
\begin{equation}
ds^2 = - a^2(r) dt^2 + \frac{dr^2}{a^2(r)} + b^2(r) \sum_{i=1}^d dx_i^2, \qquad A = A_t(r) dt,
\end{equation}
then the solution for the gauge field is
\begin{equation}
F_{tr} = q_e b^{-d} \mathrm{e}^{-2 \alpha \phi},
\end{equation}
and the other equations of motion can be reformulated as
\begin{eqnarray}
\phi'^2 + \frac{d}{2} \frac{b''}{b} &=& 0, \label{2eq5}
\\
\left( a^2 b^{2d - 2} \right)'' - 2 (d-2) a b^{d-2} b' \left( a b^{d-1} \right)'
+ 4 \Lambda b^{2d-2} &=& 0, \label{2eq6}
\\
\left( a^2 b^{d} \phi' \right)' + \alpha q_e^2 b^{-d}
\mathrm{e}^{- 2 \alpha \phi} &=& 0, \label{2eq7}
\end{eqnarray}
together with the first order constraint coming from the $rr$-component of the Einstein equation
\begin{equation} \label{2eq8}
d a a' b b' + \frac12 d (d-1) a^2 b'^2 + \Lambda b^2
= \phi'^2 a^2 b^2 - q_e^2 b^{-2d+2} \mathrm{e}^{- 2 \alpha \phi}.
\end{equation}

In order to find the near-horizon solution, we define the
new variable $w = r - r_H$ where $r_H$ denotes the radius of the horizon,
and the scaling solution should be like
\begin{equation}
\label{2eq9} a(w) = a_0 w^\gamma, \qquad b(w) = b_0 w^\beta, \qquad
\phi(w) = - k_0 \ln w,
\end{equation}
where $a_0, b_0$ and $k_0$ are constants.
After some algebra we can find the following zero-temperature solution
(fixing $b_0 = 1$ by rescaling $x_i$):
\begin{eqnarray} \label{2eq10}
&& \gamma = 1, \qquad \beta = \frac{\alpha^2}{\alpha^2 + 2 d},
\nonumber\\
&& a_0^2 = - \frac{2 \Lambda}{(d \beta + 1) (d \beta - \beta + 1)},
\quad q_e^2 = - \frac{2 \Lambda}{\alpha^2 + 2},
\quad k_0 = \frac{\alpha d^2}{2 (\alpha^2 + 2 d)}.
\end{eqnarray}
The corresponding finite-temperature solution can be easily generalized,
\begin{equation}
\label{2eq12} ds^2 = - a^2(w) f(w) dt^2 + \frac{dw^2}{a^2(w) f(w)} +
b^2(w) \sum_{i=1}^d dx_i^2, \qquad f(w) = 1 - \frac{w_0^{d\beta +
1}}{w^{d\beta + 1}},
\end{equation}
with the other fields and parameters remaining invariant.

Here are some comments on these solutions:
\begin{itemize}
\item
The zero temperature solutions with Lifshitz-like
scaling symmetry has already been obtained in~\cite{Taylor:2008tg},
and the above solutions reduce to those obtained in~\cite{Goldstein:2009cv} when $d = 2$.

\item
The near-horizon metric takes a Lifshitz-like form with anisotropic
scaling~\cite{Kachru:2008yh} whose dynamical exponent is $z = 1/\beta$.
However, it cannot be treated as the genuine gravity dual of the Lifshitz
fixed-point, since the scaling symmetry is broken by the non-trivial
dilaton.\footnote{Related work on Lifshitz black holes is listed in~\cite{others}}

\item
Following the spirit of~\cite{Goldstein:2009cv}, here we are
interested in finding asymptotically AdS$_{d+2}$ solutions,
the near horizon geometries of which are either (\ref{2eq9})
for zero temperature or (\ref{2eq12}) for finite temperature (both are exact solutions).
From~(\ref{2eq10}) we can see that the charge parameter $q_e$ is
fixed, while in the asymptotically AdS$_{d+2}$ case $q_e$ is related
to the number density in the dual field theory.

\end{itemize}

\section{Thermodynamics}
We will discuss the thermodynamics of the finite-temperature solution in this section.
First we recall the finite-temperature solution,
\begin{equation}
ds^2 = - a_0^2 w^2 f(w) dt^2 + \frac{dw^2}{a_0^2 w^2 f(w)} + w^{2 \beta}
\sum_{i=1}^d dx_i^2, \qquad f(w) = 1 - \frac{w_0^{d\beta+1}}{w^{d\beta+1}}.
\end{equation}
As $w \to \infty$, this solution reduces to the original scaling
solution. Since the scaling solution~(\ref{2eq10}) corresponds to
the near horizon of an extremal black hole, it can be expected that
the finite-temperature solution corresponds to the near-horizon
region of a near-extremal black hole.

The temperature of the black hole is given by
\begin{equation}
T = \frac{(\beta d + 1) a_0^2}{4 \pi} w_0,
\end{equation}
and the entropy density is
\begin{equation}
s = \frac14 b^d(w)\Big|_{w=w_0} = \frac14 w_0^{\beta d} \sim T^{\beta d}.
\end{equation}
For a charged black hole, the entropy density can be expressed as a
function of the temperature $T$ and the chemical potential $\mu$. Since
the dimensions of $T$ and $\mu$ are ${\rm dim} \; T = {\rm dim} \; \mu = [M]$,
the entropy density of a slightly non-extremal black hole in $(d+2)$-dimensions is
\begin{equation}
s \sim T^{\beta d} \mu^{d - \beta d}
\end{equation}
by dimensional analysis. The entropy density can also be obtained by
the standard Euclidean path integral, which gives
\begin{equation}
s = a C T^{\beta d} \mu^{d - \beta d}, \qquad C \sim L^{d}/G_{d+2}.
\end{equation}
Here $G_{d+2}$ denotes the $(d+2)$-dimensional Newton constant, and
the coefficient $a$ depends on $\alpha$ and the asymptotic value of
the dilaton $\phi_0$. Here the specific heat,
\begin{equation}
C_{v} = T \left( \frac{ds}{dT} \right)_\mu = a C \beta d T^{\beta d} \mu^{d - \beta d},
\end{equation}
is always positive. The other thermodynamical quantities can be
obtained by the entropy density via the Gibbs-Duhem relation
$s dT - dP + n d\mu = 0$. Here $P$ and $n$ are the pressure and number density.
Keeping $\mu$ fixed and performing the integration gives
\begin{equation}
P = \int s dT = \frac{a}{\beta d + 1} C \mu^{d - \beta d} T^{\beta d + 1} + P_0(\mu),
\end{equation}
where $P_0(\mu)$ is a temperature independent integration constant
which can be fixed by dimensional analysis:
\begin{equation}
P_0(\mu) = b C \mathrm{e}^{(d+1) \alpha \phi_0} \mu^{d+1}.
\end{equation}
Then from the relation
\begin{equation}
dP = a C \mu^{d - \beta d} T^{\beta d} dT + \frac{(d - \beta d)}{\beta d + 1}
a C \mu^{d - 1 - \beta d} T^{\beta d + 1} d\mu + b C (d + 1) \mathrm{e}^{(d+1)
\alpha \phi_0} \mu^d d\mu,
\end{equation}
one can identify the number density $n$ as
\begin{equation}
n = \frac{(d - \beta d)}{\beta d + 1} a C\mu^{d - 1 - \beta d}
 T^{\beta d + 1} + b C (d + 1) \mathrm{e}^{(d+1) \alpha \phi_0} \mu^d.
\end{equation}

Finally, the energy density is determined by the relation $\rho = Ts
+ \mu n - P$, which gives
\begin{equation}\label{3eq11}
\rho = \frac{d}{\beta d + 1} a C \mu^{d - \beta d} T^{\beta d + 1}
+ d b C \mathrm{e}^{(d+1) \alpha \phi_0} \mu^{d+1}.
\end{equation}
The equation of state of this near-extremal system is
\begin{equation}
P = \frac{1}{d} \rho.
\end{equation}
The susceptibility is given by
\begin{equation}
\chi \equiv \left( \frac{\partial n}{\partial \mu} \right)_T =
\frac{(d - 1 - \beta d)(d - \beta d)}{\beta d + 1} a C T^{\beta d + 1}
\mu^{d - 2 - \beta d} + d (d + 1) b C \mathrm{e}^{(d+1) \alpha \phi_0} \mu^{d - 1},
\end{equation}
which is positive when $T \ll \mu$. Notice that the first term
becomes negative when $d - 1 - \beta d < 0$, i.e. $\beta > (d-1)/d$.
As emphasized in~\cite{Goldstein:2009cv}, the formulae in this section are valid
when $T \ll \mu$. Furthermore, whether $\chi$ changes sign as the
temperature increases, signalling a phase transition, requires one to
explore beyond the regime $T \ll \mu$.

Rather than constructing the global solution which is asymptotically AdS$_{d+2}$,
we focus on some qualitative behavior of the asymptotic solution. Similar to the
four-dimensional example~\cite{Goldstein:2009cv}, the bulk solutions related by
a rescaling of coordinates should be treated as being distinct with different
chemical potential. Therefore, all solutions can be obtained by a
suitable rescaling and shift in the dilaton from a particular simple solution,
e.g. $q_e = 1$ and $\phi_0 = 0$. The rescaling is given by
\begin{equation}\label{3eq14}
r \to \lambda r, \qquad (t, x_i) \to \lambda^{-1} (t, x_i).
\end{equation}
Furthermore, from the equations of motion we can see that the metric
and $\phi - \phi_0$ only depend on $q_e^2 \mathrm{e}^{- 2 \alpha \phi_0}$.

Reconsidering the equations of motion~(\ref{2eq5})--(\ref{2eq7}) and the
constraint~(\ref{2eq8}), we can see that for an asymptotically AdS$_{d+2}$ solution,
the metric and dilaton must take the following form:
\begin{eqnarray}
a^2(r) &=& r^2 \left( 1 - e_1 \frac{\rho}{r^{d+1}} +
\frac{q_e^2 \mathrm{e}^{- 2 \alpha \phi_0}}{r^{2d}} + \cdots \right),
\nonumber\\
b^2(r) &=& r^2 \left( 1 + \cdots \right),
\nonumber\\
\phi &=& \phi_0 + \frac{\phi_1}{r^{d+1}} + \cdots,
\end{eqnarray}
where the ellipses denote terms that are subdominant at large $r$.
The parameter $\rho$ is the energy density of the black hole, and
$e_1$ is a constant depending on $L$. Under the rescaling
of~(\ref{3eq14}), the corresponding rescaling of the energy density and
charge parameter should be
\begin{equation}
\rho \to \lambda^{d+1} \rho, \qquad q_e \to \lambda^d q_e.
\end{equation}
This implies the following relation:
\begin{equation}
\rho = D_1 \left( q_e \mathrm{e}^{- \alpha \phi_0} \right)^{\frac{d+1}{d}},
\end{equation}
where $D_1$ is an $\alpha$ dependent parameter. Similarly, the chemical potential,
\begin{equation}
\mu = \int_{r_h}^{\infty} \frac{q_e}{b^d(r)} \mathrm{e}^{- 2 \alpha \phi} dr,
\end{equation}
can be determined by a similar rescaling argument as
\begin{equation}
\mu = D_2 \left( q_e \mathrm{e}^{- \alpha \phi_0} \right)^{\frac{1}{d}} \mathrm{e}^{- \alpha \phi_0},
\end{equation}
where $D_2$ is also an $\alpha$ dependent parameter. This gives
\begin{equation}
\rho = D_3 \mathrm{e}^{(d+1) \alpha \phi_0} \mu^{d+1},
\end{equation}
which agrees with the second term in~(\ref{3eq11}).

%

\section{Conductivity in zero-temperature backgrounds}
We will calculate the conductivity $\sigma$ in the $(d+2)$-dimensional extremal
black hole background, generalizing the result obtained in~\cite{Goldstein:2009cv}.
A useful formulation was proposed in~\cite{Horowitz:2009ij}, which stated that after
introducing a perturbative gauge field $A_x(r,t)$, the corresponding
equation of motion for $A_x$ could be recast in a Schr\"{o}dinger-like form
\begin{equation}\label{4eq1}
- A_{x, z z} + V(z) A_x = \omega^2 A_x,
\end{equation}
where $z$ is a redefinition of the radial variable $r$. Then by studying scattering
with ingoing boundary condition at the horizon, the conductivity is determined in
terms of the reflection coefficient
\begin{equation}
\sigma(\omega) = \frac{1 - \mathcal{R}}{1 + \mathcal{R}}.
\end{equation}
It has been pointed out in~\cite{Horowitz:2009ij} that such a
formulation could be generalized to higher-dimensional cases.

In the following we will perform the calculations in our $(d+2)$-dimensional
extremal background. Our task is still to find the equation for $A_x$ and cast
it in the form of~(\ref{4eq1}). Consider the general metric of the form
\begin{equation}\label{4eq3}
ds^2 = - g(r) \mathrm{e}^{- \chi(r)} dt^2 + \frac{dr^2}{g(r)} + r^2 \sum_{i=1}^d dx_i^2,
\end{equation}
which is extensively adopted in the discussions of holographic
superconductors~\cite{Hartnoll:2008kx}. The gauge field,
including the perturbation, is given by
\begin{equation}
A = A_t(r) dt + \tilde A_x(r) \mathrm{e}^{- i \omega t} dx.
\end{equation}
Consider the Lagrangian of the following form:
\begin{equation}
\mathcal{L} = \cdots - 2 (\nabla\phi)^2 - \frac14 f^2(\phi) F_{\mu\nu} F^{\mu\nu} + \cdots,
\end{equation}
where the gauge coupling is $f^2(\phi)$. The $t$-component of
the Maxwell equation determines the background $A_t$,
\begin{equation}\label{4prAt}
\partial_r A_t = q_e f^{-2}(\phi) r^{-d} \mathrm{e}^{-\chi/2},
\end{equation}
and the $x$-component equation,
\begin{eqnarray} \label{4eq7}
&& \omega^2 f^2(\phi) r^{d - 2} g^{-1} \mathrm{e}^{\chi/2} \tilde A_x + \partial_r \left( f^2(\phi) r^{d-2} \mathrm{e}^{-\chi/2} g \partial_r \tilde A_x \right)
\nonumber\\
&& \quad + f^2(\phi) r^{d-2} \mathrm{e}^{\chi/2} \partial_r A_t \left( \partial_r \tilde g_{tx} - \frac2{r} \tilde g_{tx} \right) = 0.
\end{eqnarray}
Notice that $g_{tx}$ should be turned on at the same order as the gauge field perturbation and we denote $g_{tx}(t, r) = \mathrm{e}^{- i \omega t} \tilde g_{tx}(r)$. Furthermore, the $rx$-component of the Einstein equations gives
\begin{equation} \label{4eq8}
\partial_r \tilde g_{tx} - \frac2{r} \tilde g_{tx} = - f^2(\phi) \tilde A_x \partial_r A_t.
\end{equation}
Substituting~(\ref{4eq8}) into~(\ref{4eq7}), we can obtain
\begin{equation}
\partial_r \left( f^2(\phi) r^{d-2} g \mathrm{e}^{-\chi/2} \partial_r \tilde A_x \right) + \omega^2 f^2(\phi) r^{d-2} g^{-1} \mathrm{e}^{\chi/2} \tilde A_x - f^4(\phi) r^{d-2} \mathrm{e}^{\chi/2} (\partial_r A_t)^2 \tilde A_x = 0,
\end{equation}
which agrees with $(3.7)$ of~\cite{Goldstein:2009cv} for $d = 2$. The background $\partial_r A_t$ is given by~(\ref{4prAt}). By taking a new coordinate $z$ and a new wavefunction $\Psi$,
\begin{equation}
\partial_z = \mathrm{e}^{-\chi/2} g \, \partial_r, \qquad \Psi = f(\phi) r^{\frac{d-2}{2}} \tilde A_x.
\end{equation}
this equation becomes a Schr\"{o}dinger equation:
\begin{equation}
- \Psi'' + V(z) \Psi = \omega^2 \Psi,
\end{equation}
where the potential is
\begin{equation}\label{4eq13}
V(z) = f^{-1}(\phi) r^{-\frac{d-2}2} \partial_z^2 \left( f(\phi) r^{\frac{d-2}2} \right) + q_e^2 f^{-2}(\phi) r^{-2d} g \mathrm{e}^{-\chi}.
\end{equation}
Here prime stands for the derivative with respect to $z$, and it agrees with $(3.15)$ of~\cite{Goldstein:2009cv} once again when $d = 2$.

\subsection{Near horizon analysis}
Recall the metric of the zero-temperature background,
\begin{equation}
ds^2 = - a_0^2 w^2 dt^2 + \frac{dw^2}{a_0^2 w^2} + w^{2 \beta} \sum_{i=1}^d dx_i^2.
\end{equation}
Comparing with~(\ref{4eq3}), we can obtain
\begin{equation}
r = w^\beta, \qquad g(r) = \beta^2 a_0^2 r^2, \qquad \mathrm{e}^{-\chi(r)} = \frac{1}{\beta^2} r^{\frac{2}{\beta} - 2}.
\end{equation}
Then
\begin{equation}\label{4eq15}
\frac{\partial r}{\partial z} = \mathrm{e}^{-\chi/2} g = \beta a_0^2 r^{\frac{1}{\beta}+1} \quad \Rightarrow \quad z = - \frac{1}{a_0^2 w}.
\end{equation}
Here the gauge coupling is
\begin{equation}\label{4eq16}
f(\phi) = 2 \exp(\alpha \phi) = \frac{2}{w^{\beta d}}.
\end{equation}
Plugging~(\ref{4eq15}) and~(\ref{4eq16}) into~(\ref{4eq13}), we obtain the following expression for $V(z)$
\begin{equation}
V_0(z) = \frac{c_0}{z^2},
\end{equation}
where
\begin{equation}
c_0 = \frac{(d+2)^2}{4} \beta^2 - \frac{d+2}{2} \beta + \frac{q_e^2}{4 a_0^2}.
\end{equation}
It can be seen that for a general $(d+2)$-dimensional extremal charged dilaton black holes, the equation of the gauge field perturbation $\tilde A_x$ can always be transformed into a Schr\"odinger equation. Furthermore, the effective potential takes a universal form $V_0(z) = c_0/z^2$, where $c_0$ is a constant which is determined by $\alpha$ and $d$.

The technique of solving the Schr\"odinger equation with specific potential is summarized in the Appendix. The ingoing mode solution is
\begin{equation}
\Psi(z) = C_0^\mathrm{(in)} \sqrt{- \frac{\pi \omega z}2} H_{\nu_0}^{(1)}(- \omega z) \sim C_0^\mathrm{(in)} \mathrm{e}^{- i (\omega z + \frac12 \nu_0 \pi + \frac12 \pi)},
\end{equation}
where $\nu_0^2 = c_0 + 1/4$.

\subsection{Asymptotic analysis}
For the asymptotic solution we have
\begin{equation}
\chi = 0, \qquad g = r^2, \qquad f = f(\phi_0),
\end{equation}
so
\begin{equation}
\frac{\partial r}{\partial z} = r^2 \quad \Rightarrow \quad z = - \frac{1}{r}.
\end{equation}
Unlike the $d = 2$ case in~\cite{Goldstein:2009cv}, a crucial difference is that the effective potential cannot be neglected at the asymptotic boundary:
\begin{equation}
V_\infty(z) = \frac{c_\infty}{z^2}, \qquad c_\infty = \frac{d (d-2)}{4}.
\end{equation}

The general solution is (refer to the Appendix for the details)
\begin{eqnarray} \label{APsi}
\Psi(z) &=& \frac{\pi}{\Gamma(\nu_\infty)} \sqrt{- \omega z} \left( C_\infty^{(1)} H^{(1)}_{\nu_\infty}(- \omega z) + C_\infty^{(2)} H^{(2)}_{\nu_\infty}(- \omega z) \right)
\nonumber\\
&\to& - i \left( C_\infty^{(1)} - C_\infty^{(2)} \right) \sqrt{- \omega z} \left( - \frac{2}{\omega z} \right)^{\nu_\infty},
\end{eqnarray}
where $\nu_\infty = \sqrt{c_\infty + 1/4} = (d-1)/2$.

\subsection{Matching}
In order to match the coefficients in the near horizon and asymptotic analysis, we should take the small $\omega$ limit to extrapolate the near horizon and asymptotic wavefunctions to an intermediate region of small $- \omega z$. From the near horizon side ($-z \gg 1$), we have
\begin{equation}
\Psi(z) = C_0^\mathrm{(in)} \sqrt{- \frac{\pi \omega z}2} H_{\nu_0}^{(1)}(- \omega z) \to (- \omega z)^{\frac12 - \nu_0},
\end{equation}
and from the asymptotic side it is just~(\ref{APsi}). The frequency
dependence can be neglected in the intermediate region. Therefore
the $\omega$-dependence of the essential combination of coefficients
can be determined:
\begin{equation} \label{CComega}
C_\infty^{(1)} - C_\infty^{(2)} \sim \omega^{\nu_\infty - \nu_0}.
\end{equation}

\subsection{Conductivity}
Next we calculate the conductivity in a general $(d+2)$-dimensional spacetime. It can be seen that the asymptotic form of $\tilde A_x$ is
\begin{equation}
\tilde A_x = \tilde A_x^{(0)} + \frac{\tilde A_x^{(1)}}{r^{d-1}},
\end{equation}
and the conductivity takes the following form:
\begin{equation}
\sigma = - i \frac{d - 1}{\omega} f^2(\phi_0) \frac{\tilde A_x^{(1)}}{\tilde A_x^{(0)}},
\end{equation}
where $f(\phi_{0})$ denotes the asymptotic value of the gauge
coupling. Therefore
\begin{equation} \label{4eq22}
(d-1) f^2(\phi_0) \left( \tilde A_x^{(1) \ast} \tilde A_x^{(0)} - \tilde A_x^{(1)} \tilde A_x^{(0)\ast} \right) = - 2 i \omega \left| \tilde A_x^{(0)} \right|^2 \, {\rm Re} \, \sigma,
\end{equation}
and the asymptotic form of $\Psi$ can be written as
\begin{equation}
\Psi = f(\phi_0) \left( \tilde A_x^{(0)} r^{\frac{d-2}{2}} + \tilde A_x^{(1)} r^{-\frac{d}{2}} \right).
\end{equation}
Then we can obtain the conserved flux at the boundary:
\begin{eqnarray}
\mathcal{F} &=& i ( \Psi^{\ast} \partial_z \Psi - \Psi \partial_z \Psi^{\ast} )
\nonumber\\
&=& i (d-1) f^2(\phi_0) \left( \tilde A_x^{(1) \ast} \tilde A_x^{(0)} - \tilde A_x^{(1)} \tilde A_x^{(0)\ast} \right) \frac{1}{r^2} \frac{\partial r}{\partial z}.
\end{eqnarray}
Substituting~(\ref{4eq22}) and noting that $\partial r/\partial z = r^2$ at the boundary, we have
\begin{equation}\label{4eq25}
\mathcal{F} = 2 \omega \left| \tilde A_x^{(0)} \right|^2 \, {\rm Re} \, \sigma.
\end{equation}

Notice that $\Psi \sim r^{d/2-1} \tilde A_x^{(0)} \sim (-z)^{1-d/2} \tilde A_x^{(0)}$, then from the results~(\ref{APsi}) and~(\ref{CComega}) we have
\begin{equation}
\tilde A_x^{(0)} = - i (2)^{\nu_\infty} \left( C_\infty^{(1)} - C_\infty^{(2)} \right) \omega^{\frac12 - \nu_\infty} \sim \omega^{\frac12 - \nu_0}.
\end{equation}
By evaluating the conserved flux at the horizon, we can easily check that
\begin{equation}
\mathcal{F} \sim \omega,
\end{equation}
and thus, combining with the result~(\ref{4eq25}), the real part of the conductivity is
\begin{equation}
{\rm Re} \, \sigma \sim\omega^{\delta}, \qquad \delta = 2 \nu_0 - 1.
\end{equation}
The exponent $\delta$ has the same expression as the $d = 2$ case in~\cite{Goldstein:2009cv}, but the value of $\nu_0$ generically depends on the spacetime dimension and also on the dilaton coupling $\alpha$.


\section{Gauss-Bonnet corrections to $\eta/s$ at finite temperature}
In this section we will discuss the Gauss-Bonnet corrections to $\eta/s$
at finite temperature in five dimensions. One remarkable progress in the
AdS/CFT correspondence is the calculation of the ratio of shear
viscosity over the entropy density in the dual gravity side. It has been
found that $\eta/s = 1/4\pi$ for a large class of CFTs with Einstein
gravity duals in the large N limit. Therefore, it was conjectured
that $1/4\pi$ is a universal lower bound for all materials, which is
the so-called Kovtun-Son-Starinets (KSS) bound~\cite{Kovtun:2004de}.
However, in~\cite{Kats:2007mq, Brigante:2007nu, Brigante:2008gz} it
was observed that in $R^{2}$ gravity such a lower bound was violated,
and a new lower bound $4/25\pi$ was proposed by considering the
causality of the dual field theory.

It was argued in~\cite{Brustein:2007jj} that the shear viscosity is
fully determined by the effective coupling of the transverse
gravitons on the horizon. This was confirmed in~\cite{Iqbal:2008by}
via the scalar membrane paradigm and in~\cite{Cai:2008ph} by
calculating the on-shell action of the transverse gravitons.
However, the full solutions were still used in the actual calculations.
Recently, $\eta/s$ with higher derivative corrections was revisited
for various examples in~\cite{Banerjee:2009ju}. They calculated
$\eta/s$ in the presence of higher order corrections by making use
of the near horizon data only. It turned out that the results agreed
with those obtained in the previous literature. An efficient method for
computing the zero frequency limit of transport coefficients in
strongly coupled field theories described holographically by higher
derivative gravity theories was proposed in~\cite{Paulos:2009yk}.

Here we calculate $\eta/s$ for black holes in five-dimensional
Gauss-Bonnet gravity. Since the charged dilaton black holes have
vanishing entropy at extremality, we shall not consider the
zero-temperature limit. We adopt the formalism proposed
in~\cite{Cai:2009zv}, where a three-dimensional effective metric
$\tilde{g}_{\mu\nu}$ was introduced and the transverse gravitons
were minimally coupled to this new effective metric. The action in
this new formalism can take a covariant form. Similar discussions on
this issue were also presented in~\cite{Banerjee:2009wg}.

Consider a tensor perturbation $h_{xy} = h_{xy}(t, u, z)$, where $u$ is
the radial coordinate in which the horizon is located at $u = 1$, and
the momentum of the perturbation points along the $z$-axis.
If the transverse gravitons can be
decoupled from other perturbations, the effective bulk action of the
transverse gravitons can be written in a general form:
\begin{equation}
S = \frac{V_{x,y}}{16\pi G_5} \left( -\frac12 \right) \int d^3x \sqrt{- \tilde g} \left[ \tilde K(u) \tilde g^{MN}
\tilde \nabla_M \tilde\phi \tilde \nabla_N \tilde\phi + m^2 \tilde\phi^2 \right],
\end{equation}
up to some total derivatives, where $\tilde\phi = h^x_y$ can be
expanded as $\tilde\phi(t,u,z) = \tilde\phi(u) \mathrm{e}^{- i \omega t + i p z}$. Here
$\tilde g_{MN}, {M, N = t, u, z}$ is a three-dimensional effective
metric, $m$ is an effective mass and $\tilde \nabla_M$ is the
covariant derivative using $\tilde g_{MN}$. Notice that $\tilde\phi$ is a
scalar in the three dimensions $t, u, z$, while it is not a scalar in
the whole five dimensions. The three-dimensional effective action
itself is general covariant, and $\tilde K(u)$ is a scalar under
general coordinate transformations. In the following we will use
$g_{\mu\nu}$ to denote the whole five-dimensional background.

The action of the transverse gravitons in momentum space can be
written explicitly as follows
\begin{equation}\label{5eq2}
S = \frac{V_{x,y}}{16 \pi G_5} \left( -\frac12 \right) \int \frac{d\omega dp}{(2\pi)^2} du
\sqrt{-\tilde g} \left[ \tilde K(u) \left( \tilde g^{uu} \tilde\phi' \tilde\phi'
+ \omega^2 \tilde g^{tt} \tilde\phi^2 + p^2 \tilde g^{zz} \tilde\phi^2 \right) + m^2 \tilde\phi^2 \right],
\end{equation}
where
\begin{eqnarray}
&& \tilde\phi(t,u,z) = \int \frac{d\omega dp}{(2\pi)^2} \, \tilde\phi(u; k) \, \mathrm{e}^{- i \omega t + i p z},
\nonumber\\
&& k = (\omega, 0, p), \qquad \tilde\phi(u; -k) = \tilde\phi^{\ast}(u; k),
\end{eqnarray}
and the prime denotes the derivative with respect to $u$.
Following~\cite{Cai:2009zv}, $\eta$ is given by
\begin{equation}
\eta = \frac1{16\pi G_5} \left[ \sqrt{\tilde g_{zz}} \, \tilde K(u) \right]_{u=1}.
\end{equation}

Next, consider a general background
\begin{equation}\label{5eq5}
ds^2 = -g(u) (1 - u) dt^2 + \frac{du^2}{h(u) (1 - u)} + \frac{r_0^2}{u^\kappa}
( dx^2 + dy^2 + dz^2 ),
\end{equation}
where $g(u)$ and $h(u)$ are regular functions at the horizon $u = 1$
and $\kappa$ is a parameter. It turns out that the effective action of
the transverse gravitons can be written in the form of~(\ref{5eq2})
with the effective three-dimensional metric
\begin{eqnarray}
\tilde{g}^{uu} &=& \left( 1 + \frac{\lambda_{\rm GB}}{2} \frac{\kappa g'_{tt} g^{uu}}{u g_{tt}} \right) g^{uu},
\\
\tilde{g}^{tt} &=& \left[ 1 + \frac{\lambda_{\rm GB}}{2} \left( \frac{\kappa g'^{uu}}{u} - \frac{(\kappa^2 + 2 \kappa) g^{uu}}{u^2} \right) \right] g^{tt},
\\
\tilde{g}^{zz} &=& \left[ 1 + \frac{\lambda_{\rm GB}}{2} \left( \frac{g'^2_{tt} g^{uu}}{g_{tt}^2} - \frac{g'_{tt} g'^{uu}}{g_{tt}} - \frac{2 g^{uu} g''_{tt}}{g_{tt}} \right) \right] g^{zz}.
\end{eqnarray}

In fact, the effective action of the transverse gravitons can also
be written as
\begin{equation}
S = \frac{1}{16\pi G_5} \left( -\frac12 \right) \int d^5x \sqrt{- g} \, \hat g^{\mu\nu} \partial_\mu \tilde\phi \partial_\nu \tilde\phi,
\end{equation}
where the new metric integrated Gauss-Bonnet correction is given by $\hat g^{\mu\nu} = \tilde g^{\mu\nu}$ for $\mu, \nu = t, u, z$ and $\hat g^{\mu\nu} = g^{\mu\nu}$ for $\mu, \nu = x, y$. Then the coupling can be computed by $\tilde K(u) = \sqrt{- g}/\sqrt{- \tilde g}$. After a straightforward calculation one can finally derive the following expression:
\begin{equation}\label{5eq10}
\frac{\eta}{s} = \frac{1}{4\pi} \left[ 1 - \frac{\kappa}{2} \lambda_{\rm GB} h(1) \right],
\end{equation}
where we have used the fact that the Bekenstein-Hawking formula
still holds in Gauss-Bonnet gravity. Notice that in order to obtain
corrections to $\eta/s$ at the leading order of $\lambda_{\rm GB}$,
it is sufficient to work in the original background~(\ref{5eq5}).

Recall the five-dimensional black hole solution
\begin{equation}
ds^2 = -a^2(w) f(w) dt^2 + \frac{dw^2}{a^2(w) f(w)} + b^2(w) (dx^2 + dy^2 + dz^2), \quad f(w) = 1 - \frac{w_0^{3\beta+1}}{w^{3\beta+1}},
\end{equation}
we can take the following coordinate transformation:
\begin{equation}
\left( \frac{w_0}{w} \right)^{3\beta + 1} = u^2,
\end{equation}
to convert the black hole metric into the form of~(\ref{5eq5}) with
\begin{eqnarray}
&& r_0 = \omega_0^\beta, \qquad \kappa = \frac{4\beta}{3\beta+1},
\nonumber\\
&& g(u) = - a_0^2 w_0^2 u^{-\frac{4}{3\beta+1}} (1 + u), \qquad h(u) = \frac{(3\beta+1)^{2}}{4} a_0^2 u^2 (1 + u).
\end{eqnarray}
Now substituting all the relevant data into~(\ref{5eq10}), we can
arrive at
\begin{equation}
\frac{\eta}{s} = \frac{1}{4\pi} \left( 1 - \frac{12 \beta}{2\beta+1} \lambda_{\rm GB} \right).
\end{equation}
Notice that $\beta \to 1$, that is, in the relativistic limit,
it reproduces the well-known result obtained in~\cite{Brigante:2007nu}.

It has been verified that in certain charged black hole backgrounds,
the charge parameter $q_e$ also contributes to the corrections to
$\eta/s$~\cite{Myers:2008yi, Ge:2008ni, Buchel:2008vz,
Cremonini:2008tw, Myers:2009ij, Cremonini:2009sy}. However, it seems
that our result does not have any dependence on $q_e$. This may be
understood as follows: in~\cite{Banerjee:2009ju} the near horizon
configuration for charged black holes contained the charge parameter
$q_e$, while here in the near horizon metric, the charge parameter
$q_e$ is fixed only by the parameter $\alpha$ after choosing
$b_0 = 1$. Then by restoring the explicit dependence of $b_0$ in
the metric, the near horizon data should contain the charge
parameter $q_e$. Therefore one can expect that the explicit
$q_e$ dependence in the corrections to $\eta/s$ might be
recovered.

\section{Summary and discussion}
In this paper we study general $(d+2)$-dimensional charged dilaton
black hole with planar symmetry obtained in~\cite{Taylor:2008tg},
generalizing the investigations in~\cite{Goldstein:2009cv}. Rather
than treating these black holes as global solutions, here we
consider them to be the near horizon solutions of a generic black
hole with AdS$_{d+2}$ asymptotic geometry. We discuss the
thermodynamics of the near-extremal black holes, and we calculate the
AC conductivity in the zero-temperature background. We find that the
AC conductivity behaves as $\omega^{\delta}$, where $\delta$ is a
constant determined by the parameter $\alpha$ in the gauge coupling
and $d$. When $d=2$, we reproduce the result obtained
in~\cite{Goldstein:2009cv}. We also calculate the Gauss-Bonnet
corrections to $\eta/s$ in a five-dimensional finite-temperature
background. The result reduces to the previously known result in the
relativistic limit. However, unlike other works studying the
higher order corrections to $\eta/s$ for charged black holes, our
result does not depend on the charge parameter $q_e$. This may be
due to the fact that the charge parameter is fixed by $\alpha$ and
$d$ after choosing a specific value for $b_{0}$, thus the near
horizon configuration does not contain information about $q_e$
explicitly. The $q_e$ dependence of the corrections to $\eta/s$
might be recovered by restoring the explicit dependence of $b_0$
in the metric.

One further generalization is to discuss the case of a dyonic black
hole, which carries both electric and magnetic charges. One can
expect that such solutions possess Lifshitz-like near horizon
geometry and an AdS$_{d+2}$ asymptotic geometry. It would be
interesting to study the thermodynamics and transport coefficients,
such as the Hall conductivity~\cite{Hartnoll:2007ai}, in the
presence of the magnetic field.

There have been several interesting papers investigating non-Fermi
liquid states in an RN-AdS black hole background~\cite{Lee:2008xf,
Liu:2009dm, Cubrovic:2009ye, Faulkner:2009wj}. The asymptotic
geometry is AdS$_{d+2}$ and the near horizon geometry contains an
AdS$_2$ part, which plays a central role in the investigations. It
would be worthwhile to generalize their considerations to the
solutions discussed here. Note that now we have a Lifshitz-like near
horizon geometry instead, and in principle we can still calculate the
corresponding correlation functions by making use of the matching
technique. We expect to study such fascinating topics in the future.

\bigskip \goodbreak
\centerline{\bf Acknowledgements}
\noindent DWP would like to thank Rene Meyer for helpful discussions.
The work of CMC was supported by the National Science Council of the
R.O.C. under the grant NSC 96-2112-M-008-006-MY3 and in part by the
National Center of Theoretical Sciences (NCTS).
The work of DWP was supported by the National Research
Foundation of Korea (NRF) grant funded by the Korea government (MEST)
through the Center for Quantum Spacetime (CQUeST) of Sogang
University with grant number 2005-0049409.

\appendix
\section{Solving the Schr\"odinger equation}
The Schr\"odinger equation with a $z^{-2}$ potential
\begin{equation}
- \Psi'' + V(z) \Psi = \omega^2 \Psi, \qquad V(z) = \frac{c}{z^2},
\end{equation}
can be transformed, by introducing a new variable (the range of $z$ is $- \infty < z < 0$)
\begin{equation}
\Psi(z) = \chi_0 \sqrt{- \omega z} \, \chi(z),
\end{equation}
to the Bessel equation:
\begin{equation}
z^2 \partial_z^2 \chi + z \partial_z \chi + ( \omega^2 z^2 - \nu^2 ) \chi = 0, \qquad \nu^2 = c + \frac14.
\end{equation}
The solutions are the Hankel functions
\begin{equation}
\chi(z) = C_1 H^{(1)}_\nu(- \omega z) + C_2 H^{(2)}_\nu(- \omega z).
\end{equation}
The approximative formulae for the Hankel functions are~\cite{AbSt64}
\begin{eqnarray}
H^{(1)}_\nu(-\omega z) &\to& - i \frac{\Gamma(\nu)}{\pi} \left( \frac{-\omega z}2 \right)^{-\nu}, \qquad -\omega z \to 0,
\\
H^{(2)}_\nu(-\omega z) &\to& i \frac{\Gamma(\nu)}{\pi} \left( \frac{-\omega z}2 \right)^{-\nu}, \qquad -\omega z \to 0,
\\
H^{(1)}_\nu(-\omega z) &\sim& \sqrt{- \frac{2}{\pi \omega z}} \; \mathrm{e}^{- i (\omega z + \frac12 \nu \pi + \frac12 \pi)}, \qquad -\omega z \sim \infty,
\\
H^{(2)}_\nu(-\omega z) &\sim& \sqrt{- \frac{2}{\pi \omega z}} \; \mathrm{e}^{i (\omega z + \frac12 \nu \pi + \frac12 \pi)}, \qquad -\omega z \sim \infty.
\end{eqnarray}


\end{document}